\documentclass[epj]{webofc}
\usepackage[varg]{txfonts}   % Web of Conferences font
\wocname{EPJ Web of Conferences}
\woctitle{CONF12}
\newcommand{\nn}{\nonumber}
\newcommand{\ket}{\,\rangle}
\newcommand{\bra}{\langle \,}
\newcommand{\mL}{\mathcal{L}}

\newcommand{\mF}{\mathcal{F}}

\newcommand{\mO}{\mathcal{O}}

\newcommand{\mT}{\mathcal{T}}
\newcommand{\mV}{\mathcal{V}}

\newcommand{\cO}{{\cal O}}

\def\bat{\begin{array}{cc}}
\begin{document}
\selectlanguage{english}
\title{Integrating out resonances in strongly-coupled electroweak scenarios~\footnote{Talk given at the XIIth Quark Confinement and the Hadron
  Spectrum, 28 August - 4 September (2016), Thessaloniki (Greece). Preprint numbers: IFIC-16-89 , FTUV-16-3011.}}
%
% subtitle (optional, strongly discouraged)

\author{Ignasi Rosell\inst{1}\fnsep\thanks{Speaker} \fnsep \thanks{\email{rosell@uchceu.es}} \and
        Antonio Pich\inst{2} \and
        Joaqu\'{\i}n Santos \inst{2} \and
        Juan Jos\'e Sanz-Cillero \inst{3}
        % etc.
}

\institute{Departamento de Matem\'aticas, F\'\i sica y Ciencias Tecnol\'ogicas, Universidad CEU Cardenal Herrera, \\ 46115 Alfara del Patriarca, Val\`encia, Spain 
\and
           Departament de F\'\i sica Te\`orica, IFIC, Universitat de Val\`encia -- CSIC, 46071 Val\`encia, Spain
\and
           Departamento de F\'\i sica Te\'orica I, Universidad Complutense de Madrid, 28040 Madrid, Spain
}

\abstract{
Accepting that there is a mass gap above the electroweak scale, the Electroweak Effective Theory (EWET) is an appropriate tool to describe this situation%, where only the Standard Model fields are taken into account
. Since the EWET couplings contain information on the unknown high-energy dynamics, we consider a generic strongly-coupled scenario of electroweak symmetry breaking, where the known particle fields are coupled to heavier states. Then, and by integrating out these heavy fields, we study the tracks of the lightest resonances into the couplings. The determination of the low-energy couplings (LECs) in terms of resonance parameters can be made more precise by considering a proper short-distance behaviour on the Lagrangian with heavy states, since the number of resonance couplings is then reduced. Notice that we adopt a generic non-linear realization of the electroweak symmetry breaking with a singlet Higgs.
}
\maketitle

\section{Motivation}
\label{motivation}

Having at hand all the recent experimental information coming from the LHC, it is accepted that the Standard Model (SM) gives a successful description of electroweak and strong interactions. Actually, pursuits of Physics beyond the SM have failed up to now, shifting to higher scales possible new degrees of freedom (dof), that is, a mass gap seems to exist between SM fields and higher scales and, consequently, effective field theories (EFTs) are appropriate to describe this regime. At low energies only the SM dof are present and the corresponding EFT, the Electroweak Effective Theory (EWET), contains a leading-order (LO) Lagrangian corresponding to the SM one and possible heavier contributions can be analyzed through the next-to-leading (NLO) Lagrangian. In other words, these higher-dimensional opperators suppressed by the corresponding powers of the new-physics scale contain fundamental information of the underlying dynamics. This is the main aim of this work~\cite{this_work,PRD}. While a direct search for these new scales is fruitless, a precise analysis of the NLO operators of the EWET is a good place to look for information of these new scales. 

The Achilles' heel of EFTs is the large number of NLO operators and their corresponding unknown low-energy couplings (LECs). %Therefore, it is fundamental to establish the assumed patterns in order to build the effective Lagrangian. 
In our case, and in order to test if the Higgs field belongs to a doublet representation, we have assumed the more general non-linear realization with a single Higgs. At high energies, and as a matter of simplification, we have considered color-singlet heavy fields with bosonic quantum numbers $J^P=0^\pm$ and $1^\pm$ that are in the first singlet or triplet representations of the electroweak group. Then, we build a general effective Lagrangian implementing the spontaneous electroweak symmetry breaking (EWSB) $SU(2)_L \otimes SU(2)_R \to SU(2)_{L+R}$ and containing the SM fields and, at high energies, the previously indicated heavier states. The scope of this project is to estimate the LECs of the EWET in terms of resonance parameters coming from high energies in order to be able to analyze possible new-physics scales.

It is interesting to stress the similarites of the EWSB and the Chiral Symmetry Breaking (ChSB) occuring in Quantum Chromodynamics (QCD), being replaced the pion decay constant $f_\pi=0.090\,$GeV by the electroweak scale $v=(\sqrt{2}G_F)^{-1/2}=246\,$GeV. Interestingly, a na\"\i ve rescaling from QCD, $M_\rho=0.77\,$GeV and $M_{a1}=1.3\,$GeV, to the electroweak sector would imply vector and axial-vector resonances of $2.1\,$TeV and $3.4\,$TeV respectively. Thus, we can make profit of our previous experience in QCD~\cite{ChPT_RChT}, where we estimated some LECs of Chiral Perturbation Theory (ChPT)~\cite{ChPT} in terms of resonance parameters by using the Resonance Chiral Theory~\cite{RChT}.

%%%%%%%%%%%%%%%%%%%%%%%%%%%%%%%%%%%%%%%%%%%%%%%%%%%%%%%%%%%%%%%
%\begin{figure}
%\begin{center}
%\includegraphics[scale=0.45]{ST-2WSR} \includegraphics[scale=0.55]{ST-1WSR}
%\caption{\small{{\bf Left}. NLO determinations of $S$ and $T$, imposing two WSRs on the $W^3B$ correlator~\cite{PRL}. The grid lines correspond to $M_V$ values from $1.5$ to $6.0$~TeV, at intervals of $0.5$~TeV, and $\kw=M_V^2/M_A^2= 0, 0.25, 0.50, 0.75, 1$. The arrows indicate the directions of growing  $M_V$ and $\kw$. The ellipses give the experimentally allowed regions at 68\% (orange), 95\% (green) and 99\% (blue) CL. {\bf Right}. Scatter plot for the 68\% CL region, in the case when only the first WSR is assumed~\cite{PRL}. The dark blue and light gray regions correspond, respectively,  to $0.2<M_V/M_A<1$ and $0.02<M_V/M_A<0.2$. We consider $M_A>M_V>0.4$~TeV in the plot.}} \label{fig.WSR}
%\end{center}
%\end{figure}
%%%%%%%%%%%%%%%%%%%%%%%%%%%%%%%%%%%%%%%%%%%%%%%%%%%%%%%%%%%%%%%

As a matter of phenomenological motivation it is interesting to quote Ref.~\cite{PRL}, a NLO analysis of the oblique electroweak observables $S$ and $T$~\cite{Peskin}. Under reasonable short-distance assumptions and by using the experimental results~\cite{exp} we checked that there is room for this kind of strongly-coupled scenarios once the resonance masses appeared at the TeV scale and the $WW$ coupling of the Higgs is close to the SM value, $\kappa_W\simeq 1$.%, as shown in figure~\ref{fig.WSR}. As it can be observed, the presence of the second Weinberg Sum Rule (WSR)~\cite{WSR} is not fundamental for the result.

These proceedings are organized in the following way. In Sect.~\ref{Lagrangian} we construct both effective Lagrangians: at low energies (with only the SM fields) and at high energies (with SM fields and resonances), Sect.~\ref{EWET} and \ref{resonances} respectively. We have looked at different formalisms for the massive spin-$1$ fields, {\it i.e.}, the Proca and the antisymmetric formalisms, in order to prove their full equivalence once proper high-energy information is taken into account, see Sect.~\ref{PandA}. In Sect.~\ref{LECs} the heavy resonances are integrated out to be able to estimate the LECs of the EWET. The possibility of considering short-distance constraints in order to reduce the number of resonance parameters, increasing then the predictive power of this analysis, is addressed in Sect.~\ref{short-distance}.

\section{Building the Lagrangian}
\label{Lagrangian}

\subsection{Low energies: SM fields}
\label{EWET}

The EWET is built by considering the most general Lagrangian containing the SM dof ($W^\pm$ and $Z$ gauge bosons, fermions, electroweak Goldstones and the Higgs $h$), satisfying the SM symmetries and following the pattern ot the EWSB: $G\equiv SU(2)_L\otimes SU(2)_R\rightarrow H\equiv SU(2)_{L+R}$. We follow the notation of Ref.~\cite{this_work,PRD}:
\begin{itemize}
\item The Goldstone fields are parametrized through the canonical $G/H$ coset representative~\cite{RChT} $u(\varphi) =\exp{(\frac{i}{2}\,\vec\sigma\vec\varphi/v)}$, so under chiral transformations $g\equiv (g_L^{\phantom{\dagger}},g_R^{\phantom{\dagger}})\in G$, $u(\varphi)\to g_L^{\phantom{\dagger}} u(\varphi) g_h^\dagger(\varphi,g) = g_h^{\phantom{\dagger}}(\varphi,g) u(\varphi) g^\dagger_R$, with $g_h^{\phantom{\dagger}}(\varphi,g)\equiv g_h^{\phantom{\dagger}}\in H$. For convenience, we consider $U=u^2\to g_L^{\phantom{\dagger}} U g^\dagger_R$ and $u_\mu = i\, u\, (D_\mu U)^\dagger u = u_\mu^\dagger\to g_h^{\phantom{\dagger}} u_\mu g_h^{\dagger}$. 
\item The covariant derivative $D_\mu U = \partial_\mu U - i \hat W_\mu U + i U \hat B_\mu$ couples the Goldstones to external $SU(2)_{L,R}$ gauge sources, making the Lagrangian formally invariant under local $G$ transformations. The identification with the SM gauge fields, $\hat W_\mu = -\frac{g}{2}\,\vec\sigma\vec W_\mu$ and $\hat B_\mu = -\frac{g'}{2}\,\sigma_3 B_\mu$, breaks explicitly the $SU(2)_R$ symmetry while preserving the $SU(2)_L\otimes U(1)_Y$ SM symmetry. 
\item The left and right field-strength tensors have been re-written in terms of $f_\pm^{\mu\nu}\equiv u^\dagger \hat W^{\mu\nu} u \pm u\, \hat B^{\mu\nu} u^\dagger$, which transform as triplets under $G$: $f_\pm^{\mu\nu} \to g_h^{\phantom{\dagger}} f_\pm^{\mu\nu} g_h^{\dagger}$. 
\item The fermions transform under $G$ like $\psi_L  \rightarrow  g_X  g_L  \psi_L $ and $\psi_R \rightarrow g_X g_R  \psi_R$ with $g_X \in U(1)_X$. In order to construct the EWET operators, it is convenient to introduce the covariant fermion doublet fields $\xi_{L} \equiv u_{L}^\dagger  \psi_{L} = u^\dagger  \psi_L$ and $\xi_{R} \equiv  u_{R}^\dagger  \psi_{R}   = u \psi_R$, which transform with $g_h$ instead of $g_{L,R}$: $\xi_{L,R}\rightarrow   g_X  g_h \xi_{L,R}$. In the Lagrangian we introduce the fermions by considering the following bilinears: $(J_S)_{mn}\equiv \bar{\xi}_n \xi_m$, $(J_P)_{mn}\equiv i\,   \bar{\xi}_n \gamma_5 \xi_m$, $(J_V^\mu)_{mn}\equiv \bar{\xi}_n\gamma^\mu \xi_m$, $(J_A^\mu)_{mn} \equiv \bar{\xi}_n\gamma^\mu\gamma_5 \xi_m$ and $(J_T^{\mu\nu})_{mn} \equiv \bar{\xi}_n\sigma^{\mu\nu} \xi_m$.
\item We use $\mathcal{T}$ to introduce an explicit breaking of custodial symmetry, $\mathcal{T}= u \mathcal{T}_R u^\dagger \rightarrow\quad g_h^{\phantom{\dagger}} \mathcal{T} g_h^\dagger$, being $\mathcal{T}_R$ the right-handed spurion $\mathcal{T}_R\rightarrow g_R^{\phantom{\dagger}} \mathcal{T}_R g_R^\dagger$ and making the identification $\mathcal{T}_R  = -g'\frac{\sigma_3}{2}$.
\end{itemize}

%%%%%%%%%%%%%%%%%%%%%%%%%%%%% Table %%%%%%%%%%%%%%%%%%%%%%%%%%%%%%%
\begin{table}[t!]
\begin{center}
\renewcommand{\arraystretch}{1.4}
\caption{\small
Bosonic operators of the $\cO(p^4)$ 
EWET Lagrangian~\cite{this_work,PRD}. $\mO_i$ ($\widetilde\mO_i$) denote $P$-even (odd) structures.}
\label{tab:bosonic-EWET}
\begin{tabular}{|c||c|c|}
\hline
\multicolumn{1}{|c||}{$i$} &
\multicolumn{1}{|c|}{${\cal O}_i$} &
\multicolumn{1}{|c|}{$\widetilde{\cal O}_i$}  \\
\hline
\hline
$1$  &
$\frac{1}{4}\bra {f}_+^{\mu\nu} {f}_{+\, \mu\nu}
- {f}_-^{\mu\nu} {f}_{-\, \mu\nu}\ket$
&  $\frac{i}{2} \bra {f}_-^{\mu\nu} [u_\mu, u_\nu] \ket$
\\ [1ex]
\hline
$2$  &
$ \frac{1}{2} \bra {f}_+^{\mu\nu} {f}_{+\, \mu\nu}
+ {f}_-^{\mu\nu} {f}_{-\, \mu\nu}\ket$
& $\bra {f}_+^{\mu\nu} {f}_{-\, \mu\nu} \ket $
\\ [1ex]
\hline
$3$  &
$\frac{i}{2}  \bra {f}_+^{\mu\nu} [u_\mu, u_\nu] \ket$
&  $\frac{(\partial_\mu h)}{v}\,\bra f_+^{\mu\nu}u_\nu \ket$
\\ [1ex]
\hline
$4$  &
$\bra u_\mu u_\nu\ket \, \bra u^\mu u^\nu\ket $
& ---
\\ [1ex]
\hline
$5$  &
$  \bra u_\mu u^\mu\ket^2$
& ---
\\ [1ex]
\hline
$6$ &
$\frac{(\partial_\mu h)(\partial^\mu h)}{v^2}\,\bra u_\nu u^\nu \ket$
& ---
\\ [1ex]
\hline
$7$  &
$\frac{(\partial_\mu h)(\partial_\nu h)}{v^2} \,\bra u^\mu u^\nu \ket$
& ---
\\ [1ex]
\hline
$8$ &
$\frac{(\partial_\mu h)(\partial^\mu h)(\partial_\nu h)(\partial^\nu h)}{v^4}$
& ---
\\ [1ex]
\hline
$9$ &
$\frac{(\partial_\mu h)}{v}\,\bra f_-^{\mu\nu}u_\nu \ket$
&  ---
\\ [1ex]
\hline
$10$ & $\langle \mT u_\mu\rangle^2$  & ---
\\ [1ex]
\hline
$11$ & $ \hat{X}_{\mu\nu} \hat{X}^{\mu\nu}$ & ---
\\ [1ex]
\hline
\multicolumn{3}{c}{}
\end{tabular}
\end{center}
%\vspace*{-1.cm}
\end{table}
%%%%%%%%%%%%%%%%%%%%%%%%%%%%%%%%%%%%%%%%%%%%%%%%%%%%%%%%%%%%%%%%%%%%%

The effective Lagrangian is organized as a low-energy expansion in powers of momenta, %$\mathcal{L}_{\mathrm{EWET}} = \sum_{\hat d\ge 2}\, \mathcal{L}_{\mathrm{EWET}}^{(\hat d)}$, 
\begin{equation}
\mathcal{L}_{\mathrm{EWET}} \,=\, \sum_{\hat d\ge 2}\, \mathcal{L}_{\mathrm{EWET}}^{(\hat d)}\,, 
\end{equation}where the operators cannot be simply ordered according to their canonical dimensions and one must use instead the so-called chiral dimension $\hat d$ which reflects their infrared behaviour at low momenta~\cite{ChPT}. Quantum loops are renormalized order by order in this low-energy expansion. The power-counting rules can be summarized as: $h/v  \sim \mathcal{O}\left(p^0\right)$; $u_\mu,\, \partial_\mu$ and $\mathcal{T} \sim \mathcal{O}\left(p\right)$; $ f_{\pm\, \mu\nu},\, \hat{X}_{\mu\nu},\, J_{S,P}$ and $J_{V,A}^{\mu}  \sim \mathcal{O}\left(p^2\right)$. It is interesting to spotlight two features related to this power counting:
\begin{enumerate}
\item Assuming that the SM fermions couple weakly to the strong sector we assign an $\cO(p^2)$ to fermion bilinears. Considering  na\"\i vely a chiral analysis an $\cO( p)$ would have been assigned.
\item Considering the phenomenology, and contrary to the first papers studying the Higgsless EWET~\cite{Longhitano}, we assign an $\cO( p)$ to the explicit breaking of custodial symmetry %(incorporated in our notation through $\mT$).
\end{enumerate}

As it has been pointed out previously the LO Lagrangian corresponds to the SM one. The NLO Lagrangian~\cite{this_work,PRD,Longhitano,Cata} can be split in different pieces,
\begin{equation}
\mL_{\mathrm{EWET}}^{(4)} \, = \, \sum_{i=1}^{11} \mF_i \; \mO_i  + \sum_{i=1}^{3}\widetilde\mF_i \; \widetilde \mO_i + \sum_{i=1}^{7}  \mF_i^{\psi^2}\; \mO_i^{\psi^2}  + \sum_{i=1}^{3}   \widetilde\mF_i^{\psi^2}\; \widetilde \mO_i^{\psi^2} + \sum_{i=1}^{10}\mF_i^{\psi^4}\; \mO_i^{\psi^4} + \sum_{i=1}^{2}\widetilde\mF_i^{\psi^4}\; \widetilde \mO_i^{\psi^4} \, ,
\label{EWET-Lagrangian}
\end{equation}
where the operators have been separated considering their $P$ nature (without or with tilde for $P$-even and $P$-odd operators) and the presence of fermions. In tables~\ref{tab:bosonic-EWET} and \ref{tab:fermion-EWET} we show all the operators. Note that the different LECs are not simple constants, since they can be multiplied by an arbitray polinomial of $h$~\cite{Grinstein:2007iv}. Different NLO calculations with the EWET can be found in the literature, see Refs.~\cite{NLO_altres}.

%%%%%%%%%%%%%%%%%%%%%%%%%%%%  Table  %%%%%%%%%%%%%%%%%%%%%%%%%%%%%%%
\begin{table}[!t] %[tbh]
\begin{center}
\renewcommand{\arraystretch}{1.4}
\caption{\small
Fermion operators of the $\cO(p^4)$ EWET Lagrangian~\cite{this_work}. $\mO_i^{\psi^2,\psi^4}$ ($\widetilde\mO_i^{\psi^2,\psi^4}$) denote $P$-even (odd) structures.}
\label{tab:fermion-EWET}

\begin{tabular}{|c||c|c||c|c|}
\hline
$i$ & ${\cal O}^{\psi^2}_i$ & $\widetilde{\cal O}^{\psi^2}_i$
& ${\cal O}^{\psi^4}_i$ & $\widetilde{\cal O}^{\psi^4}_i$
\\ \hline\hline
1  & $\bra J_S \ket \bra u_\mu u^\mu \ket$  &
$\bra J_T^{\mu \nu} f_{- \,\mu\nu} \ket $
& $\bra J_{S} J_{S} \ket $ &  $\bra J_V^\mu J_{A,\mu}^{\phantom{\mu}}\ket $
\\  [1ex] \hline
2 &  $ i \,  \bra J_T^{\mu\nu} \left[ u_\mu, u_\nu \right] \ket$
& $\frac{\partial_\mu h}{v} \, \bra u_\nu J^{\mu\nu}_T \ket $
& $\bra J_{P} J_{P} \ket $ &  $\bra J_V^\mu\ket \bra J_{A,\mu}^{\phantom{\mu}}\ket $
\\  [1ex] \hline
3 & $\bra J_T^{\mu \nu} f_{+ \,\mu\nu} \ket $ & $\bra J_V^\mu \ket \bra u_\mu \mathcal{T} \ket $ & $\bra J_{S} \ket \bra  J_{S} \ket $ & ---
\\  [1ex] \hline
4 & $\hat{X}_{\mu\nu} \bra J_T^{\mu \nu} \ket $ & ---
& $\bra J_{P} \ket \bra  J_{P} \ket $ & ---
\\ [1ex] \hline
5 & $\frac{\partial_\mu h}{v} \, \bra u^\mu J_P \ket $ & ---
&  $\bra J_V^\mu J_{V,\mu}^{\phantom{\mu}}\ket $ & ---
\\ [1ex] \hline
6 & $\bra J_A^\mu \ket \bra u_\mu \mathcal{T} \ket $ & --- &
 $\bra J_A^\mu J_{A,\mu}^{\phantom{\mu}}\ket $  & ---
\\ [1ex] \hline
7 &  $\frac{(\partial_\mu h) (\partial^\mu h)}{v^2} \bra J_S\ket $           %%%---
& --- &
 $\bra J_V^\mu\ket \bra J_{V,\mu}^{\phantom{\mu}}\ket $  & ---
\\ [1ex] \hline
8 & --- & --- &
 $\bra J_A^\mu\ket \bra J_{A,\mu}^{\phantom{\mu}}\ket $  & ---
\\ [1ex] \hline
9 & --- & --- &
$\bra J^{\mu\nu}_{T} J_{T\,\mu\nu}^{\phantom{\mu}} \ket $ & ---
\\ [1ex] \hline
10 & --- & --- &
$\bra J^{\mu\nu}_{T} \ket \bra J_{T\,\mu\nu}^{\phantom{\mu}} \ket $ & ---
\\ [1ex] \hline
\end{tabular}
\end{center}
\end{table}
%%%%%%%%%%%%%%%%%%%%%%%%%%%%%%%%%%%%%%%%%%%%%%%%%%%%%%%%%%%%%%%%%%%%%

\subsection{High energies: SM fields and resonances}
\label{resonances}

At higher energies we have to consider also resonance fields: scalar ($S$), pseudoscalar ($P$), vector ($V$) and axial-vector ($A$) resonances in our case. We consider generic massive states, transforming under $G$ as $SU(2)_{L+R}$ triplets ($R= \sigma^a R^a/\sqrt{2}$) or singlets ($R_{1}$): $R \rightarrow g_h^{\phantom{\dagger}} R g_h^\dagger$ and $R_{1}\rightarrow R_{1}$ respectively. We can split the Lagrangian in terms which contain explicitly resonances, $\mL_{\rm R}$, and terms which do not contain resonances, $\mL_{\text{non-R}}$,
\begin{equation}
\mL_{\mathrm{RT}} \,=\, \mL_{\rm R}[R,\chi,\psi] + \mL_{\text{non-R}}[\chi,\psi]\, .
\label{RT-Lagrangian}
\end{equation}
Note that the second piece is formally identical to the EWET Lagrangian of (\ref{EWET-Lagrangian}), but with different couplings, because it describes the interactions of a different EFT, valid at the resonance mass scale.

The spin-$0$ Lagrangian reads
\begin{eqnarray}
\mL_R & =& \frac{1}{2}\bra \nabla^\mu R\,   \nabla_\mu R  - M_R^2\, R^2\ket  + \bra R\, \chi_R^{\phantom{\mu}}\ket \hskip 2.5cm (R=S,\, P)\, , \nn\\
\mL_{R_1} & =&  \frac{1}{2}  \left(  \partial^\mu R_1\,  \partial_\mu R_1  - M_{R_1}^2\, R_1^2 \right)   + R_1\, \chi_{R_1}^{\phantom{\mu}} \hskip 2.5cm (R_1=S_1,\, P_1)\, ,
\label{SP-Lagrangian}
\end{eqnarray}
being the interactions given by~\cite{this_work,PRD}
\begin{align}
\chi_S^{\phantom{\mu}}  &=  c_1^S\;  J_S \, , &
\chi_P^{\phantom{\mu}}  &=  c_1^P\; J_P  +  d_P\; \frac{(\partial_\mu h)}{v}\, u^\mu \, , \nn \\  
\chi_{S_1}^{\phantom{\mu}}  &=  \lambda_{hS_1} \, v \; h^2  + \frac{c_{d}}{\sqrt{2}}\; \bra u_\mu u^\mu \ket \; +\; \frac{c_1^{S_1}}{\sqrt{2}}\; \bra J_S \ket  \, ,&
\chi_{P_1}^{\phantom{\mu}} &=  \frac{c_1^{P_1}}{\sqrt{2}}\;  \bra  J_P\ket \, .
\label{SP-interactions}
\end{align}
Note that we have considered only terms linear in the heavy resonances and of $\cO(p^2)$. As before, we follow the notation of Ref.~\cite{this_work,PRD}.

In the case of spin-$1$ fields there is freedom in the representation to be chosen. We have considered both the Proca and the antisymmetric formalism, since we want to prove their equivalence. In order to avoid any misunderstanding, we use $\hat{R}$ and the superindex $(P )$ in the case of the Proca formalism, whereas we use $R$ and the superindex $(A )$ in the antisymmetric case. Including again only interactions linear in the four-vector fields, the relevant chiral Lagrangians in the Proca formalism take the form:
\begin{eqnarray}
\mL_{\hat R}^{(P)} & =& - \frac{1}{4}\,\bra \hat R_{\mu \nu}\, \hat R^{\mu \nu} -  2\,M_{R}^2\, \hat R_\mu \hat R^\mu \ket +  \bra \hat R_{\mu}\, \hat \chi^{\mu}_{\hat R} + \hat R_{\mu \nu}\, \hat \chi_{\hat R}^{\mu \nu} \ket \qquad \qquad\quad (\hat R=\hat V,\, \hat A)\, , \nn\\
\mL_{\hat R_1}^{(P)} & =& - \frac{1}{4} \left( \hat R_{1\, \mu \nu}\, \hat R_1^{\mu \nu} -  2\,M_{R_1}^2\, \hat R_{1\,\mu} \hat R_1^\mu \right)  +  \hat R_{1\,\mu}\, \hat \chi^{\mu}_{\hat R_1}  + \hat R_{1\,\mu \nu}\, \hat \chi_{\hat R_1}^{\mu \nu} \qquad\quad (\hat R_1 = \hat V_1,\, \hat A_1)\, ,
\label{VAProca-Lagrangian}
\end{eqnarray}
where $R_{\mu \nu}  = \nabla_\mu \hat R_\nu - \nabla_\nu \hat R_\mu$ and $R_{1\,\mu \nu}  = \partial_\mu \hat R_{1\,\nu} - \partial_\nu  \hat R_{1\,\mu}$.
%\begin{equation}
%\hat R_{\mu \nu} \, =\, \nabla_\mu \hat R_\nu - \nabla_\nu \hat R_\mu\, ,
%\qquad\qquad\qquad
%\hat R_{1\,\mu \nu} \, =\, \partial_\mu \hat R_{1\,\nu} - \partial_\nu  \hat R_{1\,\mu} \, .
%\label{Rhat_mn}
%\end{equation}
The interactions are given at $\cO(p^2)$ by~\cite{this_work}:
\begin{eqnarray}
\hat \chi_{\hat V}^{\mu \nu}  & =& \frac{f_{\hat V}}{2 \sqrt 2} \, f_{+}^{\mu\nu} +  \frac{i\, g_{\hat V}}{2 \sqrt 2} \, [u^\mu,u^\nu]  +  \frac{\widetilde f_{\hat V}}{2\sqrt 2} \, f_{-}^{\mu\nu} + \frac{  \widetilde{\lambda}_1^{h\hat{V}} }{\sqrt{2}}\;\left[ (\partial^\mu h)\, u^\nu-(\partial^\nu h)\, u^\mu \right] + c_{0}^{\hat{V}} J_T^{\mu\nu} \, , \nn\\[10pt]
\hat \chi_{\hat A}^{\mu \nu}  & =& \frac{f_{\hat A}}{2 \sqrt 2} \, f_{-}^{\mu\nu} + \frac{ \lambda_1^{h\hat{A}} }{\sqrt{2}}\;\left[ (\partial^\mu h)\, u^\nu-(\partial^\nu h)\, u^\mu \right]  +  \frac{\widetilde f_{\hat A}}{2\sqrt 2} \, f_{+}^{\mu\nu} +  \frac{i\, \widetilde g_{\hat A}}{2 \sqrt 2} \, [u^\mu,u^\nu] + \widetilde{c}_0^{\hat{A}} J_T^{\mu\nu} \, , \nn\\[10pt]
\hat \chi_{\hat V_1}^{\mu\nu}  & =& f_{\hat{V}_1} X^{\mu\nu}+ \frac{c_0^{{\hat{V}}_1} }{\sqrt{2}} \bra J_T^{\mu\nu}\ket \, , \qquad \qquad \qquad \qquad \hat \chi_{\hat A_1}^{\mu\nu}  = \widetilde{f}_{\hat{A}_1} X^{\mu\nu} + \frac{\widetilde c_0^{{\hat{A}}_1}}{\sqrt{2}} \bra J_T^{\mu\nu}\ket\, ,  \nn \\ [10pt]
\hat \chi_{\hat V}^\mu  & =& c_1^{\hat{V}}\, J^\mu_V  + \widetilde c_1^{\hat{V}}\, J^\mu_A\, , \qquad\qquad\qquad\qquad\qquad \;\;\; \hat \chi_{\hat A}^\mu = c_1^{\hat{A}}\, J^\mu_A  + \widetilde c_1^{\hat{A}}\, J^\mu_V\, , \nn\\[10pt]
 \hat \chi_{\hat V_1}^\mu  & =& \widetilde{c}_{\mathcal{T}}^{\hat{V}_1} \bra u^\mu \mathcal{T} \ket + \frac{c_1^{{\hat{V}}_1}}{\sqrt{2}} \, \bra J^\mu_V\ket  + \frac{   \widetilde c_1^{{\hat{V}}_1}  }{\sqrt{2}} \, \bra J^\mu_A\ket\, ,\quad 
 \hat \chi_{\hat A_1}^\mu  \,=\, c_{\mathcal{T}}^{\hat{A}_1} \bra u^\mu \mathcal{T} \ket +  \frac{  c_1^{{\hat{A}}_1}  }{\sqrt{2}} \, \bra J^\mu_A\ket  + \frac{  \widetilde c_1^{{\hat{A}}_1}  }{\sqrt{2}} \, \bra J^\mu_V\ket \, .
\label{VAProca-interactions}
\end{eqnarray}

If the antisymmetric formalism is chosen to describe the spin-$1$ fields, the resonance Lagrangian reads:
\begin{eqnarray}
\mL_R^{(A)} & =& - \frac{1}{2}\bra \nabla^\lambda R_{\lambda\mu} \,  \nabla_\sigma R^{\sigma \mu} - \frac{1}{2} M_R^2\, R_{\mu\nu} R^{\mu\nu} \ket \; +\; \bra R_{\mu\nu} \chi^{\mu\nu}_R\ket \hskip 1.5cm (R=V,\, A)\, , \nn\\
\mL_{R_1}^{(A)} & =&   -   \frac{1}{2} \left( \partial^\lambda R_{1\, \lambda\mu}^{\phantom{\mu}} \,  \partial_\sigma R_1^{\sigma \mu} - \frac{1}{2} M_{R_1}^2\, R_{1\, \mu\nu}^{\phantom{\mu}} R_1^{\mu\nu} \right) + R_{1\, \mu\nu}^{\phantom{\mu}}\, \chi^{\mu\nu}_{R_1} \hskip 1.2cm (R_1=V_1,\, A_1)\, .\quad
\label{VAAntisim-Lagrangian}
\end{eqnarray}
The interactions are given now by~\cite{this_work,PRD}:
\begin{eqnarray}
\chi_V^{\mu\nu}  & =& \frac{F_V}{2\sqrt{2}}\;  f_+^{\mu\nu} + \frac{i\, G_V}{2\sqrt{2}}\; [u^\mu, u^\nu]  + \frac{\widetilde{F}_V }{2\sqrt{2}}\; f_-^{\mu\nu}  + \frac{ \widetilde{\lambda}_1^{hV} }{\sqrt{2}}\;\left[
(\partial^\mu h)\, u^\nu-(\partial^\nu h)\, u^\mu \right] \nn \\
&& \qquad + \,C_0^V J_T^{\mu\nu} + \frac{C_{1}^V}{2} \left(  \nabla^\mu J_V^\nu   -   \nabla^\nu J_V^\mu\right)  +  \frac{\widetilde{C}_{1}^V}{2} \left(  \nabla^\mu J_A^\nu - \nabla^\nu J_A^\mu \right) \,,   \nonumber \\
\chi_A^{\mu\nu}  & =& \frac{F_A}{2\sqrt{2}}\;  f_-^{\mu\nu}  + \frac{ \lambda_1^{hA} }{\sqrt{2}}\;\left[ (\partial^\mu h)\, u^\nu-(\partial^\nu h)\, u^\mu \right]+ \frac{\widetilde{F}_A}{2\sqrt{2}}\; f_+^{\mu\nu} +
\frac{i\, \widetilde{G}_A}{2\sqrt{2}}\; [u^{\mu}, u^{\nu} ]\, , \nn \\
&& \qquad +\, \widetilde{C}_0^A J_T^{\mu\nu} + \frac{C_{1}^A }{2} \left( \nabla^\mu J_A^\nu   -   \nabla^\nu J_A^\mu \right) + \frac{\widetilde{C}_{1}^A}{2} \left( \nabla^\mu J_V^\nu - \nabla^\nu J_V^\mu \right)\nonumber \\
\chi_{V_1}^{\mu\nu}  & = & F_{V_1}\; X^{\mu\nu} + \frac{\widetilde{C}_{\mathcal{T}}^{V_1}}{2} \left( \partial^\mu \bra u^\nu \mathcal{T} \ket - \partial^\nu \bra u^\mu \mathcal{T} \ket \right) \nn \\
&& \qquad + \,\frac{C_0^{V_1}}{\sqrt{2}} \bra  J_T^{\mu\nu} \ket + \frac{C_{1}^{V_1 }}{2\sqrt{2}} \, \bra \partial^\mu J_V^\nu -  \partial^\nu J_V^\mu \ket + \frac{\widetilde{C}_{ 1}^{V_1}}{2\sqrt{2}}\, \bra \partial^\mu J_A^\nu - \partial^\nu J_A^\mu \ket \,, \nonumber \\
\chi_{A_1}^{\mu\nu}  & = &\widetilde{F}_{A_1}\; X^{\mu\nu}+ \frac{C_{\mathcal{T}}^{A_1}}{2} \left( \partial^\mu \bra u^\nu \mathcal{T} \ket - \partial^\nu \bra u^\mu \mathcal{T} \ket \right) \nn \\
&& \qquad + \,\frac{\widetilde{C}_0^{A_1}}{\sqrt{2}} \bra  J_T^{\mu\nu} \ket + \frac{C_{1}^{A_1 }}{2\sqrt{2}} \, \bra \partial^\mu  J_A^\nu -  \partial^\nu J_A^\mu \ket + \frac{\widetilde{C}_{ 1}^{A_1}}{2\sqrt{2}}\, \bra\partial^\mu J_V^\nu - \partial^\nu J_V^\mu \ket  \,,
\label{VAAntisim-interactions}
\end{eqnarray}
where in the first line of every resonance contribution we show the purely bosonic pieces, while in the second one fermion contributions appear.

\subsection{Equivalence of Proca and antisymmetric formalism}
\label{PandA}

The equivalence of both formalisms can be demonstrated through a change of variables in the corresponding path integral~\cite{equivalence} and this yields the following set of relations between resonance parameters in both formalisms~\cite{this_work}:
\begin{align}
F_R\,=&\; f_{\hat{R}}\, M_R\, , \qquad\; &
G_R\,=&\; g_{\hat{R}}\, M_R\, , \qquad\; & \lambda_1^{hR} \, =&\; \lambda_1^{h\hat{R}}\, M_R\, , \qquad\; &
C_0^R \,=&\; c_0^{\hat{R}}\,  M_R\, , \nn\\
\widetilde{F}_R \,=&\; \widetilde{f}_{\hat{R}}\, M_R\, , & 
 \widetilde{G}_R \,=&\; \widetilde{g}_{\hat{R}}\, M_R\, , &
\widetilde{\lambda}_1^{hR} \, =&\; \widetilde{\lambda}_1^{h\hat{R}}\, M_R \, ,&
\widetilde{C}_0^R \,=&\; \widetilde{c}_0^{\hat{R}}\, M_R\, , \nn\\
C_{\mathcal{T}}^R \,=&\; c_{\mathcal{T}}^{\hat{R}} /M_R\, , &
\widetilde{C}_{\mathcal{T}}^R \,=&\; \widetilde{c}_{\mathcal{T}}^{\hat{R}} /M_R\, , &
C_1^R \,=&\; c_1^{\hat{R}} /M_R\, ,  &
\widetilde{C}_1^R \,=&\; \widetilde{c}_1^{\hat{R}} /M_R\, .\;
\label{eq.Proca-Antisim-relations}
\end{align}
By using (\ref{eq.Proca-Antisim-relations}) both formalisms must give the same predictions for the LECs of the EWET%, as we can study by considering specific Green functions
. For instance, we show here the expressions ot the two-Goldstone vector form factor in both formalisms~\cite{this_work},
\begin{eqnarray}
\mathbb{F}^\mV_{\varphi\varphi}(s) & =& 
\left\{ \bat 1 +\displaystyle\frac{f_{\hat{V}}\, g_{\hat{V}}}{  v^2  }\,\frac{ s^2 }{M_V^2-s} + \frac{\widetilde{f}_{\hat{A}}\,\widetilde{g}_{\hat{A}}   }{  v^2 }\,\frac{  s^2   }{M_A^2-s} - 2\, \mF_3^{\,\mathrm{SDP}}\,\frac{s}{v^2} & \qquad\quad\mbox{\small (SDET-P)} \, , \nn \\[10pt]
1 +\displaystyle\frac{F_V\,G_V}{v^2}\,\frac{s}{M_V^2-s} +\frac{\widetilde{F}_A\,\widetilde{G}_A}{v^2}\,\frac{s}{M_A^2-s}    - 2\, \mF_3^{\,\mathrm{SDA}}\,\frac{s}{v^2} & \qquad\quad\mbox{\small  (SDET-A)}\, ,  \end{array} \right. \quad  \\ \vspace{-1cm}
\end{eqnarray}
where SDET-P and SDET-A refer to the short-distance effective theories by using the Proca and the antisymmetric formalism respectively. In the same way, $\mF_3^{\,\mathrm{SDP}}$ and $\mF_3^{\,\mathrm{SDA}}$ are the corresponding non-resonant $\cO(p^4)$ coupling in both formalisms. Requiring that the vector form factor must vanish at high energies we get the following conditions:
\begin{equation}
\mF_3^{\,\mathrm{SDP}} \,=\, -\frac{f_{\hat{V}}\, g_{\hat{V}}}{2} - \frac{\widetilde{f}_{\hat{A}}\,\widetilde{g}_{\hat{A}} }{2}\,, \qquad \qquad \qquad \mF_3^{\,\mathrm{SDA}}\,=\,0\,. \label{constraint}
\end{equation}
This result is very interesting, since once (\ref{constraint}) is used, both formalisms give the same prediction for the LEC of the EWET, $\mF_3$, but the game is different in each formalism. While in the Proca formalism the LEC is determined by the non-resonant local term (without any contribution coming from the resonance exchange), in the antisymmetric formalism the LEC is saturated by the resonance exchange (without any contribution coming from the non-resonant local term). Note that in order to prove the equivalence of both predictions we have needed to assume a well-behaved form factor. A general analysis can be summarized in the following way~\cite{this_work}:
\begin{enumerate}
\item EWET LECs with resonance contributions coming from $\hat{\chi}^{\mu}_{\hat{R}}$ of (\ref{VAProca-interactions}) do not contain non-resonant local contributions, so then the Proca formalism is the best choice.
\item EWET LECs with resonance contributions coming from $\chi^{\mu\nu}_R$ of (\ref{VAAntisim-interactions}) do not contain non-resonant local contributions, so then the antisymmetric formalism is the best choice.
\end{enumerate}
Note that at $\cO(p^4)$ there are not resonance contributions coming from $\hat{\chi}^{\mu\nu}_{\hat{R}}$ of (\ref{VAProca-interactions}) and there are not $\cO(p^4)$ LECs with contributions coming from $\chi^{\mu\nu}_R$ and $\hat{\chi}^{\mu}_{\hat{R}}$ at the same time.

\section{Estimation of the LECs}
\label{LECs}

%%%%%%%%%%%%%%%%%%%%%%%%%%% Table %%%%%%%%%%%%%%%%%%%%%%%%%%%
\begin{table}[tb]  %%%b]
{\renewcommand{\arraystretch}{1.7}
\begin{center}
\caption{\small Prediction of purely bosonic $\cO(p^4)$ LECs from heavy resonance exchange~\cite{this_work}.}
%Prediction of $\Delta\mL_R^{\cO(p^4)}$ couplings from heavy resonance exchange to the bosonic $\cO(p^4)$ EWET Lagrangian.}
\label{tab:bLECs-final}
\begin{tabular}{|c||c|c|}
\hline
$i$ &   $\Delta\mF_i$ &  $\Delta\widetilde{\mF}_i$  \\  \hline \hline
1 &  $- \frac{F_V^2-\widetilde{F}_V^2}{4M_V^2}
+ \frac{F_A^2-\widetilde{F}_A^2}{4M_A^2} $
&
 $- \frac{\widetilde{F}_VG_V}{2M_V^2}
- \frac{F_A\widetilde{G}_A}{2M_A^2}$
\\ \hline
2 &
 $- \frac{F_V^2+{\widetilde{F}_V}^2}{8M_V^2}
- \frac{F_A^2+{\widetilde{F}_A}^2}{8M_A^2}$
&
 $- \frac{F_V \widetilde{F}_V}{4M_V^2}
- \frac{F_A \widetilde{F}_A}{4M_A^2}$
\\ \hline
3 &
$-  \frac{F_VG_V}{2M_V^2} - \frac{\widetilde{F}_A\widetilde{G}_A}{2M_A^2}$
&
 $- \frac{F_V \widetilde{\lambda}_1^{hV} v}{M_V^2}
 - \frac{\widetilde{F}_A \lambda_1^{hA} v}{M_A^2}$
\\ \hline
4 &
 $\frac{G_V^2}{4M_V^2} + \frac{{\widetilde{G}_A}^2}{4M_A^2} $
& ---
\\ \hline
5 &
$        \frac{c_{d}^2}{4M_{S_1}^2}
-\frac{G_V^2}{4M_V^2} - \frac{{\widetilde{G}_A}^2}{4M_A^2} $
& ---
\\ \hline
6 &
 $ - \frac{\widetilde{\lambda}_1^{hV\,\, 2}v^2}{M_V^2}
- \frac{\lambda_1^{hA\,\, 2}v^2}{M_A^2}$
& ---
\\ \hline
7 & $        \frac{ d_P^2}{2 M_P^2}
+ \frac{\lambda_1^{hA\,\, 2}v^2}{M_A^2}
+  \frac{\widetilde{\lambda}_1^{hV\,\, 2}v^2}{M_V^2}$
& ---
\\ \hline
8 & 0 & ---
\\ \hline
9 &  $  - \frac{F_A \lambda_1^{hA} v}{M_A^2}
- \frac{\widetilde{F}_V \widetilde{\lambda}_1^{hV} v}{M_V^2}$
& ---
\\ \hline
10 &  $-\frac{(\widetilde{c}_{\mathcal{T}}^{\hat{V}_1})^2}{2M_{V_1}^2}-\frac{(c_{\mathcal{T}}^{\hat{A}_1})^2}{2M_{A_1}^2}$ & ---
\\ \hline
11 &  $- \frac{F_{V_1}^2}{M_{V_1}^2} - \frac{\widetilde{F}_{A_1}^2}{M_{A_1}^2} $  & ---
\\ \hline
\end{tabular}
\end{center}
}
\end{table}
%%%%%%%%%%%%%%%%%%%%%%%%%%%%%%%%%%%%%%%%%%%%%%%%%%%%%%%%%%%%%%%%

%%%%%%%%%%%%%%%%%%%%%%%%%%%%  Table  %%%%%%%%%%%%%%%%%%%%%%%%%%%%%%%
\begin{table}[!t]
\begin{center}
\renewcommand{\arraystretch}{1.7}
\caption{\small Prediction of two-fermion $\cO(p^4)$ LECs from heavy resonance exchange~\cite{this_work}.}
%Final prediction of $\Delta\mL_R^{\cO(p^4)}$ couplings from heavy resonance exchange to the two-fermion $\cO(p^4)$ EWET Lagrangian.}
 \label{tab:2fLECs-final}
\begin{tabular}{|c||c|c|}
\hline
$i$ &  $\Delta\mF^{\psi^2}_i$ &  $\Delta\widetilde{\mF}^{\psi^2}_i$
\\  \hline\hline
1
& $\frac{c_d c^{S_1}_1}{2 M_{S_1}^2}$
& $-\frac{\widetilde{F}_V C_0^V}{\sqrt{2}M_V^2}\! -\! \frac{F_A \widetilde{C}_0^A}{\sqrt{2}M_A^2} $
\\[1ex] \hline
2
&  $-\frac{G_V C_0^V}{\sqrt{2}M_V^2} \!-\! \frac{\widetilde{G}_A \widetilde{C}_0^A}{\sqrt{2}M_A^2} $
&\! \!\!$-\frac{2\sqrt{2}v\widetilde{\lambda}_1^{hV}C_0^V}{M_V^2} \!-\! \frac{2\sqrt{2}v\lambda_1^{hA}\widetilde{C}_0^A}{M_A^2}$
\\[1ex] \hline
3
& $-\frac{F_V C_0^V}{\sqrt{2}M_V^2} \!-\! \frac{\widetilde{F}_A \widetilde{C}_0^A}{\sqrt{2}M_A^2} $
& $-\frac{\widetilde{c}_{\mathcal{T}}^{\hat{V}_1} c_{1}^{\hat{V}_1}    }{\sqrt{2} M_{V_1}^2} - \frac{c_{\mathcal{T}}^{\hat{A}_1} \widetilde{c}_{1}^{\hat{A}_1}    }{\sqrt{2} M_{A_1}^2}$
\\[1ex] \hline
4 & $-\frac{\sqrt{2}F_{V_1} C_0^{V_1}}{M_{V_1}^2} \!- \!\frac{\sqrt{2}\widetilde{F}_{A_1} \widetilde{C}_0^{A_1}}{M_{A_1}^2} $ & ---
\\[1ex] \hline
5 & $\frac{d_P c^{P}_1}{M_{P}^2}$ & ---
\\[1ex] \hline
6 & $-\frac{\widetilde{c}_{\mathcal{T}}^{\hat{V}_1} \widetilde{c}_{1}^{\hat{V}_1}    }{\sqrt{2} M_{V_1}^2} - \frac{c_{\mathcal{T}}^{\hat{A}_1} c_{1}^{\hat{A}_1}    }{\sqrt{2} M_{A_1}^2}$
& ---
\\[1ex] \hline
7 & $0$ & ---
\\[1ex] \hline
\end{tabular}
\end{center}
\end{table}
%%%%%%%%%%%%%%%%%%%%%%%%%%%%%%%%%%%%%%%%%%%%%%%%%%%%%%%%%%%%%%%%%%%%

The EWET LECs of the low-energy effective theory in (\ref{EWET-Lagrangian}) can be estimated in terms of resonance parameters of the high-energy effective theory in (\ref{RT-Lagrangian}) by integrating out the heavy fields, and once the preceding comments about the non-resonant local contributions are taken into account. The results are given in Tables~\ref{tab:bLECs-final}, \ref{tab:2fLECs-final} and \ref{tab:4fLECs-final} for the purely bosonic, two-fermion and four-fermion $\cO(p^4)$ LECs, respectively~\cite{this_work}. A few interesting results can be extracted from these tables~\cite{this_work}:
\begin{enumerate}
\item A non-zero $P$-odd LEC indicates a spin-1 particle with both $P$-odd and $P$-even couplings.
\item A non-zero value of any of the LECs
$\mF_{1\text{-}4,6,9\text{-}11}$, $\mF^{\psi^2}_{2\text{-}4,6}$ and $\mF^{\psi^4}_{5\text{-}10}$ indicates spin 1.
\item A non-zero value for $\mF^{\psi^2}_1$ ($\mF^{\psi^4}_1$) signals a singlet (triplet) scalar.
\item A non-zero value for $\mF_5^{\psi^2}$ or $\mF^{\psi^4}_2$ is a signal of a triplet pseudoscalar.
\item $\mF^{\psi^4}_3$ ($\mF^{\psi^4}_4$) indicates a scalar (pseudoscalar) boson.
\item The custodial-breaking LEC $\mF^{\psi^2}_6$ ($\widetilde\mF^{\psi^2}_3$) manifests a singlet $P$-odd (even) vector  or $P$-even (odd) axial-vector coupling preserving custodial symmetry, combined with a custodial-breaking
$P$-odd (odd) vector  or $P$-even (even) axial-vector coupling.
\item A non-zero value of $\mF_4 + \mF_5$ ($\mF_6 + \mF_7$) indicates a singlet scalar (triplet pseudoscalar).
\item A non-zero value $\mF_{10}$ ($\mF_{11}$) indicates a singlet $P$-odd (even) vector or $P$-even (odd) axial-vector coupling.
\item $\mF^{\psi^4}_{5,9}$ ($\mF^{\psi^4}_6$) manifest a triplet $P$-even (odd) vector  or $P$-odd (even) axial-vector coupling.
\item $\widetilde\mF_{1\text{-}3}$, $\widetilde\mF^{\psi^2}_{1,2}$ and $\widetilde\mF^{\psi^4}_1$ signal a triplet spin-1 particle.
\item A non-zero value of $\mF^{\psi^4}_1 + 2\,\mF^{\psi^4}_3$  ($\mF^{\psi^4}_2 + 2\,\mF^{\psi^4}_4$) indicates a singlet scalar (pseudoscalar).
\end{enumerate}

%%%%%%%%%%%%%%%%%%%%%%%%%%%%  Table  %%%%%%%%%%%%%%%%%%%%%%%%%%%%%%%
\begin{table}[!t]
\begin{center}
\renewcommand{\arraystretch}{1.7}
\caption{\small Prediction of four-fermion $\cO(p^4)$ LECs from heavy resonance exchange~\cite{this_work}.}
%Final prediction of $\Delta\mL_R^{\cO(p^4)}$ couplings from heavy resonance exchange to the four-fermion $\cO(p^4)$ EWET Lagrangian.}
\label{tab:4fLECs-final}
\begin{tabular}{|c||c|c|}
\hline
$i$
& $\Delta\mF_i^{\psi^4}$
& $\Delta\widetilde{\mF}_i^{\psi^4}$
\\ \hline\hline
1
&
$\frac{(c_1^S)^2}{2M_S^2} $
&
$ -
\frac{c_1^{\hat{V}} \widetilde{c}_1^{\hat{V}} }{M_V^2}
-\frac{ c_1^{\hat{A}}\widetilde{c}_1^{\hat{A}}}{M_A^2}$
\\ \hline
2
&
$\frac{(c_1^P)^2}{2M_P^2}$
&
$ \frac{c_1^{\hat{V}}\widetilde{c}_1^{\hat{V}}}{2M_V^2}
+\frac{c_1^{\hat{A}}\widetilde{c}_1^{\hat{A}}}{2M_A^2}
   -\frac{c_1^{\hat{V}_1}\widetilde{c}_1^{\hat{V}_1}}{2M_{V_1}^2} -\frac{c_1^{\hat{A}_1}\widetilde{c}_1^{\hat{A}_1}}{2M_{A_1}^2}
$
\\ \hline
3 & $-\frac{(c_1^S)^2}{4M_S^2}+\frac{(c_1^{S_1})^2}{4M_{S_1}^2}$  & ---
\\ \hline
4 & $-\frac{(c_1^P)^2}{4M_P^2}+\frac{(c_1^{P_1})^2}{4M_{P_1}^2}$  & ---
\\ \hline
5
& $ -\frac{(c_1^{\hat{V}})^2}{2M_V^2}
-\frac{(\widetilde{c}_1^{\hat{A}})^2}{2M_A^2}$
& ---
\\ \hline
6
& $- \frac{(\widetilde{c}_1^{\hat{V}})^2}{2M_V^2}
-\frac{({c}_1^{\hat{A}})^2}{2M_A^2} $
& ---
\\ \hline
7
& $\frac{({c}_1^{\hat{V}})^2}{4M_V^2}
+\frac{(\widetilde{c}_1^{\hat{A}})^2}{4M_A^2}
   -\frac{({c}_1^{\hat{V}_1})^2}{4M_{V_1}^2}  -\frac{(\widetilde{c}_1^{\hat{A}_1})^2}{4M_{A_1}^2} $
& ---
\\ \hline
8
& $ \frac{(\widetilde{c}_1^{\hat{V}})^2}{4M_V^2}
+\frac{({c}_1^{\hat{A}})^2}{4M_A^2}
   -\frac{(\widetilde{c}_1^{\hat{V}_1})^2}{4M_{V_1}^2}  -\frac{({c}_1^{\hat{A}_1})^2}{4M_{A_1}^2} $
& ---
\\ \hline
9 &
$-\frac{(C_0^V)^2}{M_V^2}-\frac{(\widetilde{C}_0^A)^2}{M_A^2}$
& ---
\\ \hline
10 &
$\frac{(C_0^V)^2}{2M_V^2}\!-\!\frac{(C_0^{V_1})^2}{2M_{V_1}^2}
\!+\!\frac{(\widetilde{C}_0^A)^2}{2M_A^2}\!-\!\frac{(\widetilde{C}_0^{A_1})^2}{2M_{A_1}^2}$
& ---
\\ \hline
\end{tabular}
\end{center}
\end{table}
%%%%%%%%%%%%%%%%%%%%%%%%%%%%%%%%%%%%%%%%%%%%%%%%%%%%%%%%%%%%%%%%%%%%%

\section{Short-distance constraints and the purely bosonic sector}
\label{short-distance}

As it can be observed in Tables~\ref{tab:bLECs-final}, \ref{tab:2fLECs-final} and \ref{tab:4fLECs-final}, the LECs are predicted in terms of many unknown resonance parameters, so it would be interesting to study possible high-energy constraints in order to reduce the number of unknown parameters and, consequently, to increase the predictive power of these results. Following this idea, and as a first approach, we have studied the prediction of $P$-even purely bosonic $\cO(p^4)$ LECs from $P$-even heavy resonance exchange (without considering operators with $\mT$ or $\hat{X}_{\mu\nu}$)~\cite{PRD}, once the following short-distance constraints are taken into consideration: the two-Goldstone vector form factor vanishes at infinite momentum transfer, the Higgs-Goldstone axial form factor vanishes at large energies and two Weinberg Sum Rules on the $W^3B$ correlator~\cite{WSR}. Then, the predictions are given in terms of only a few resonance parameters, as it is shown in Table~\ref{tab:bLECs-final2}~\cite{PRD}: the first column correponds to the $P$-even contributions of the first column of Table~\ref{tab:bLECs-final}, while in the second column the aforementioned high-energy constraints have been used. %More powerful phenomenlogical conclusions can be extracted from these results, as it can be observed in figures~\ref{fig:F1} and \ref{fig:F2}. Note, for instance, the strong contraints for $\mF_{1,3,4,6,9}$ and the possibilites given by the sums $\mF_4+\mF_5$ and $\mF_6+\mF_7$ to look for, respectively, scalar and pseudoscalar contributions. 
It is convenient to stress that, and contrary to the QCD case, the properties of the underlying theory are not well known, since one is working beyond the SM.

%%%%%%%%%%%%%%%%%%%%%%%%%%% Table %%%%%%%%%%%%%%%%%%%%%%%%%%%
\begin{table}[tb]  %%%b]
{\renewcommand{\arraystretch}{1.7}
\begin{center}
\caption{\small Prediction of purely $P$-even purely bosonic $\cO(p^4)$ LECs from $P$-even heavy resonance exchange~\cite{PRD}. The right columm includes short-distance constraints.}
\label{tab:bLECs-final2}
\begin{tabular}{|c||c|c|}
\hline
$i$ &   $\Delta\mF_i$ &  $\Delta\mF_i$  \\  \hline \hline
1 &  $\frac{F_A^2}{4M_A^2}- \frac{F_V^2}{4M_V^2}$ & $-\frac{v^2}{4}\,\left(\frac{1}{M_V^2}+\frac{1}{M_A^2}\right)$ \\ \hline
2 & $-\frac{F_A^2}{8M_A^2}- \frac{F_V^2}{8M_V^2}$ & $-\frac{v^2 (M_V^4+M_A^4)}{8 M_V^2M_A^2 (M_A^2-M_V^2)}$ \\ \hline
3 & $-  \frac{F_VG_V}{2M_V^2}$ & $-\frac{v^2}{2M_V^2}$ \\ \hline
4 & $\frac{G_V^2}{4 M_V^2}$ & $\frac{(M_A^2-M_V^2) v^2}{4 M_V^2 M_A^2}$\\ \hline
5 & $\frac{c_{d}^2}{4M_{S_1}^2} -\frac{G_V^2}{4M_V^2}$ & $\frac{c_{d}^2}{4M_{S_1}^2} -\frac{(M_A^2-M_V^2) v^2}{4 M_V^2 M_A^2}$ \\ \hline
6 & $-\frac{(\lambda_1^{hA})^2v^2}{M_A^2}$ =& $-\frac{M_V^2 (M_A^2-M_V^2) v^2}{M_A^6}$ \\ \hline
7 & $\frac{d_P^2}{2 M_P^2}+ \frac{(\lambda_1^{hA})^2v^2}{M_A^2}$ & $\frac{d_P^2}{2 M_P^2}+\frac{M_V^2 (M_A^2-M_V^2) v^2}{M_A^6}$ \\ \hline
8 & $0$ & $0$ \\ \hline
9 &  $- \frac{F_A \lambda_1^{hA} v}{M_A^2}$ & $-\frac{M_V^2 v^2}{M_A^4}$ \\ \hline
\end{tabular}
\end{center}
}
\end{table}
%%%%%%%%%%%%%%%%%%%%%%%%%%%%%%%%%%%%%%%%%%%%%%%%%%%%%%%%%%%%%%%%

\begin{acknowledgement}

We wish to thank the organizers of the conference for the pleasant conference. This work has been supported in part by the Spanish Government and ERDF funds from the European Commission (FPA2013-44773-P, FPA2014-53631-C2-1-P); by the Spanish Centro de Excelencia Severo Ochoa Programme (SEV-2012-0249, SEV-2014-0398); by the Generalitat Valenciana (PrometeoII/2013/007); by the Universidad CEU Cardenal Herrera and Banco Santander (PRCEU-UCH CON-15/03, INDI15/08) and by La Caixa (Ph.D. grant for Spanish universities).

\end{acknowledgement}

\end{document}